\documentstyle[epsf,amsfonts,prb,aps]{revtex}
\begin{document}
\draft
\preprint{\vbox{\hbox{\large G\"oteborg ITP 97-01}
}}

\title{Properties of a Luttinger Liquid with Boundaries \\
at Finite Temperature and Size}

\author{Ann E. Mattsson\cite{address}, Sebastian Eggert, 
and Henrik Johannesson}

\address{Institute of Theoretical Physics,
Chalmers University of Technology and G\"oteborg University,
S-412 96 G\"oteborg, Sweden}


\maketitle

\begin{abstract}
\widetext\leftskip=0.10753\textwidth \rightskip\leftskip
We use bosonization methods to calculate the exact finite-temperature 
single-electron Green's function of a spinful Luttinger liquid confined by 
open boundaries. The corresponding local spectral density is constructed and 
analyzed in detail. The interplay between boundary, finite-size and thermal 
effects are shown to dramatically influence the low-energy properties of 
the system. In particular, the well-known zero-temperature critical 
behavior in the bulk always crosses over to a boundary dominated regime 
in the vicinity of the Fermi level. Thermal fluctuations cause an enhanced 
depletion of spectral weight for small energies $\omega$, with the spectral 
density scaling as $\omega^2$ for $\omega$ much less than the temperature.
Consequences for photoemission experiments are discussed. 
\end{abstract}


\section{Introduction}
In the last decade there has been an enormous interest in metallic phases 
of matter which are {\em not} Fermi liquids. A paradigm for these is the 
{\em Luttinger liquid}, describing the low-energy, long-wave length 
limit of gapless electron systems in one dimension (1D)\cite{Haldane,LP}. 
The Luttinger liquid satisfies 
Luttinger's theorem, but the interaction wipes out the quasi-particle 
pole of the electron propagator, with the disjoint Fermi surface 
(consisting of two Fermi ``points'' $\pm k_F$) supporting only collective 
charge- and spin density excitations. These excitations are dynamically 
independent, effectively leading to a spatial separation of the charge 
and spin of an electron added to the system. As a result, spectral
properties and dynamical correlations
are quite different from those of a Fermi liquid\cite{Voit}.

The generic low-energy properties of a gapless interacting 1D electron 
system are well understood\cite{Haldane,Voit} 
in terms of universal power-laws of
correlation functions and spectral functions.
The behavior of such so-called {\em Luttinger liquids} is
coded in the Tomonaga-Luttinger model\cite{TomLut}, 
much in the same way as 
the normal state of 3D interacting electrons is patterned on  
the free Fermi gas {\em (Landau Fermi liquid theory)}. However, it 
is only in the last few years that laboratory technology has advanced 
to the point that the notion of a Luttinger liquid 
can be confronted with experiments. 
Indeed, the ability to manufacture true 1D quantum 
wires \cite{Tarucha,yacoby}, as well as the development of high-precision 
spectroscopic techniques for probing quasi-1D materials \cite{DardelHwu}, have 
provided strong impetus for investigating Luttinger liquid 
physics in realistic contexts.
Added motivation comes from the realization that the edge excitations of 
the fractional quantum Hall effect can be described in terms of a {\it chiral}
Luttinger liquid\cite{Wen}. Also, 
some of the ``non-Fermi liquid scenarios'' for the normal state properties 
of the cuprate superconductors draw heavily from Luttinger liquid theory, 
suggesting 
possible extensions to higher dimensions \cite{Anderson}.

Whereas Luttinger liquid theory has been successfully used to predict   
fractional quantum Hall edge state transport 
\cite{KF,Moon,Milliken}, its applicability to the more traditional 
realm of quasi-1D materials remains controversial. This class of 
materials contains organic conductors such as 
tetrathiafulvalene-tetracyanoquinodimethane (TTF-TCNQ) and the    
tetramethyltetraselenafulvalene (TMTSF)$_2$X Bechgaard salts with 
$\rm X =  ClO_4, \ PF_6,\ $ etc.   All available information suggests 
that the physics of these compounds is dominated by strong 
electron correlations and pronounced one-dimensionality \cite{Voit,Jerome}. 
Yet, data from NMR \cite{Wzietek} and photoemission spectroscopy  
\cite{DardelHwu} interpreted within a conventional 
Luttinger liquid framework seem to imply 
single-electron correlations governed by an exponent 
much larger than can be provided from any realistic lattice 
Hamiltonian (of which the Tomonaga-Luttinger model would be the effective 
low-energy continuum theory). This poses a major problem for the
modeling of these materials, and its resolution remains a challenge
for the theorist.  

One of the basic quantities to consider in this context is the
single-particle {\em spectral density},  
as this is the object that determines the outcome 
of a photoemission experiment. Surprisingly, the full calculation 
of the spectral density of an ideal, infinite volume 
Luttinger liquid was only recently performed 
\cite{MedenSchonhammerVoit}, expanding upon earlier results by Suzumura 
\cite{Suzumura}. 
In the present paper we go a step further and investigate the {\em local, 
finite-temperature spectral density} of a 
Luttinger liquid confined by {\em open boundaries} (simulating a  
reflecting barrier or edge potential). As photoemission 
spectroscopy measurements are highly sensitive to boundary effects 
(with the photoelectrons traveling only a  
short distance, of the order of a few lattice spacings from the surface 
of the sample), it is crucial to incorporate an analysis of nontrivial 
boundary conditions. Also, 
spectroscopy on high-mobility quantum wires doped with artificial 
impurities (``antidots'')\cite{Kirczenov} may soon be 
within experimental reach, adding yet another reason to study this 
problem: at sufficiently low temperatures a potential scatterer is 
expected to act essentially as a reflecting barrier\cite{KF}.  

In the present paper we
calculate the exact finite-temperature single-electron
Green's function and the associated local spectral density 
of a confined spinful Luttinger liquid. 
The combined effects of 
{\em boundaries} and {\em finite volume} will be shown to strongly modify the 
well-known bulk Luttinger liquid spectral density.
In particular, the presence of boundaries causes a novel scaling behavior 
for energies close to the Fermi level, which produce a
depletion of spectral weight significantly larger than that for a bulk 
system, as we have reported for a semi-infinite 
system at zero temperature before\cite{OurPRL}.
In addition, thermal fluctuations deplete 
the levels further and give rise to a universal $\omega^2$ scaling of the 
spectral density for sufficiently small frequencies $\omega$, which is 
analogous to the temperature effects in ``bulk'' systems\cite{nakamura}. 
Particular attention is given to the nontrivial zero mode contribution 
to the correlation function from fluctuations in the spin and charge 
quantum numbers, which turn out to violate the spin-charge separation (i.e.
the partition function cannot be split into separate spin and charge factors).
Interestingly, the confined system is sensitive to the ratio 
between the effective velocities for the collective charge- and spin 
excitations (which in turn depend on the effective electron-electron 
coupling): As the velocity ratio locks into a rational value the spectrum 
separates into distinct peaks of equal spacing. 
The spacing between the peaks becomes dense for irrational values of the 
velocity ratios, leading to
a quasi-continuous spectrum for large $\omega$. This hints at a 
resonant interference between standing waves of charge and spin for 
special values of the electron-electron coupling. 

The paper is organized as follows: In the next section we introduce
an extended version of the Tomonaga-Luttinger model and review its 
bosonization in the presence of 
open boundaries. In Sec.~III the full finite-temperature Green's
function in a bounded domain is derived, and the resulting local
spectral density is extracted and analyzed in Sec.~IV. 
Section~V contains a brief
discussion of possible consequences for photoemission experiments,
as well as some concluding remarks.

\section{1D Electrons in the Presence of Open Boundaries: Bosonization}

As our model we take an extension of the 
Tomonaga-Luttinger Hamiltonian\cite{TomLut},
 which describes spinful Fermions in one-dimension
with a (repulsive) local interaction.   All electron-electron
interactions  must conserve spin and charge, so that we use
the most general gapless Hamiltonian density that is invariant under
the corresponding SU(2) and U(1) symmetries,
\begin{eqnarray}
{\cal H}   \ = &     v_F&\left[
  \psi^\dagger_{L,\sigma} i {d \over dx} \psi^{}_{L,\sigma} 
 - \psi^\dagger_{R,\sigma} i {d \over dx} \psi^{}_{R,\sigma}\right] \nonumber \\
  & + &    g_{2 \perp}  J_{L}^{\sigma}   J_{R}^{-\sigma}   
  \ + \    g_{2 \parallel}  J_{L}^{\sigma}   J_{R}^{\sigma}   
  \ + \   g_{4 \perp}  (J_{L}^{\sigma}   J_{L}^{-\sigma} + 
J_{R}^{\sigma}   J_{R}^{-\sigma})
   \ + \ g_{4 \parallel}  (J_{L}^{\sigma}   J_{L}^{\sigma} + 
J_{R}^{\sigma}   J_{R}^{\sigma}) 
  \label{general}\\
  & +  &   g_{1 \perp}  \psi^\dagger_{L,\sigma}  \psi^{}_{R,\sigma} 
  \psi^\dagger_{R,-\sigma}  \psi^{}_{L,-\sigma} , 
\nonumber
\end{eqnarray}
where we have used the traditional ``g-ology'' scheme to index the couplings 
\cite{Solyom}. The chiral Fermion currents are defined as
$J_{L/R}^\sigma \equiv :\psi_{L/R,\sigma}^\dagger\psi_{L/R,\sigma}^{}:$, 
and
$\psi_{L/R,\sigma}(x)$ are the left/right moving components
of the electron field $\Psi_{\sigma}(x)$ expanded about the
Fermi points $\pm k_F$,
\begin{equation}
\label{psileftright}
\Psi^{\mathop{\phantom{\dagger}}}_{\sigma}(x) = 
e^{-ik_Fx}\psi^{\mathop{\phantom{\dagger}}}_{L,\sigma}(x) 
+ e^{ik_Fx}\psi^{\mathop{\phantom{\dagger}}}_{R,\sigma}(x) .
\label{decomposition}
\end{equation}
This expansion is valid as long as the lattice spacing is much smaller
than all length scales we want to consider. Therefore, 
the energy range around the Fermi-surface is limited to a
region where a linear approximation of the spectrum is justified.
This is often conveniently illustrated by a cut-off parameter, but 
more accurately one should take higher order operators in the
Hamiltonian into account which result in corrections to the linear
spectrum which are of higher orders in $|k - k_F|$.  These perturbations 
have non-universal coefficients which depend on the detailed interactions 
of the underlying lattice Hamiltonian. The appropriate value of the
cut-off parameter (i.e. the range of validity) is then determined by
the momentum scale at which those corrections become so large that
the perturbation series no longer converges.  Generically we can only
roughly estimate the range of validity to be about one order
of magnitude less than the band-width.

The first term in Eq.~(\ref{general})
is that of free relativistic fermions, while $g_2$ and $g_4$
describe forward electron-electron scattering.  We have also explicitly
included a backward scattering term $g_{1 \perp}$.
The coupling constants depend on the microscopic 
parameters of the underlying lattice model and $v_F$ is the Fermi velocity.
Normal ordering is carried out w.r.t. the
filled Dirac sea, and we sum over repeated spin indices.

Eq.~(\ref{general}) defines a ``standard model'' for low-energy electrons in a  
1D metallic phase, and is easily derived from the Hubbard
Hamiltonian 
\begin{equation}
\label{Hubbard}
 H_H = - t \sum_i \left( c^{\dagger}_{i,\sigma} 
 c^{\mathop{\phantom{\dagger}}}_{i+1,\sigma} + 
 c^{\dagger}_{i+1,\sigma} c^{\mathop{\phantom{\dagger}}}_{i,\sigma} \right)  
 + U \sum_i n^{\mathop{\phantom{\dagger}}}_{i,\uparrow} 
 n^{\mathop{\phantom{\dagger}}}_{i,\downarrow} \ , \ \ U > 0.
 \end{equation}
In the weak coupling limit $U \ll t$, we can treat $U$ as a perturbation,
and the  tight-binding band  
$\epsilon(k) = -2t \cos ak$ may be linearized around the 
Fermi points $\pm k_F = \pm n_e\pi/2a$, 
$n_e$ being the electron density and 
$a$ the lattice spacing. The electron operators are replaced by
the chiral fields $\psi^{\mathop{\phantom{\dagger}}}_{L/R,\sigma} (x)$
in the continuum limit 
\begin{equation}
\label{cleftright}
c^{\mathop{\phantom{\dagger}}}_{n,\sigma}/\sqrt{a} \approx 
e^{-ik_Fna} \psi^{\mathop{\phantom{\dagger}}}_{L,\sigma} (na)
+ e^{ik_Fna} \psi^{\mathop{\phantom{\dagger}}}_{R,\sigma} (na) \, ,
\end{equation}
As a result, the Hubbard model
gets mapped onto the Hamiltonian density in Eq.~(\ref{general})
with coupling constants $g_{1 \perp} =
g_{2 \perp} = 2g_{4 \perp} = Ua$ and $g_{2 \parallel} = g_{4 \parallel} =0$.
The Umklapp term $e^{-i 4 k_F x}\psi^\dagger_{L,\sigma}  
\psi^{}_{R,\sigma} \psi^\dagger_{L,-\sigma}  \psi^{}_{R,-\sigma} +h.c.$ 
is also generated, but does not contribute
away from half-filling
$(n_e \neq 1,\ k_F \neq \pi/2a)$ due to rapid phase oscillations. 
For this case we are left 
with the theory in (\ref{general}), with $v_F = 2at \sin (k_F a)$. 
It is important to emphasize that the Hamiltonian in (\ref{general}) 
faithfully represents the low-energy sector of the 
Hubbard model also for strong on-site repulsion $U$\cite{Schulz}.
However, when $U$ is not small, the procedure above fails to identify 
the proper values of the model parameters, and instead these have to be 
inferred from the exact {\em Bethe Ansatz} solution of the Hubbard 
model\cite{Schulz,LiebWu}.

The Hamiltonian (\ref{general}) is conveniently bosonized \cite{Stone} by 
introducing charge and spin currents and the corresponding
bosons $\phi_c$ and $\phi_s$ with 
conjugate momenta $\Pi_c$ and $\Pi_s$, respectively
\begin{mathletters} \label{JRLcsdef}
\begin{eqnarray} 
J_L^{c/s} & \equiv & \frac{1}{\sqrt{2}} \left( J_L^\uparrow \pm
J_L^\downarrow \right) = \frac{1}{\sqrt{4 \pi}} \left( {\partial_x \phi_{c/s}} +
\Pi_{c/s} \right) , \label{JLcsdef}\\
J_R^{c/s} & \equiv & \frac{1}{\sqrt{2}} \left( J_R^\uparrow \pm
J_R^\downarrow \right) = \frac{1}{\sqrt{4 \pi}} \left( {\partial_x \phi_{c/s}} -
\Pi_{c/s} \right). \label{JRcsdef}
\end{eqnarray}
\end{mathletters}
The resulting theory describes separate spin and charge
excitations moving with velocities (to lowest order in the coupling constants)
\begin{equation}
v_c = { v_F + {g_{4 \parallel} \over \pi} } + {g_{4 \perp} \over \pi}
\ \ \ \ \ 
v_s = { v_F +{g_{4 \parallel} \over \pi} } - {g_{4 \perp} \over \pi} \ \ ,
\label{vnu}
\end{equation} 
where $v_c > v_s$ for repulsive interactions
$g_{4 \perp} > 0$. The Hamiltonian becomes
\begin{eqnarray}
{\cal H}  & =  & \sum_{\nu=s,c}\left\{    \frac{v_\nu}{2}
\left[({\partial_x \phi_\nu})^2 + \Pi_\nu^2\right]
\ + \ \frac{g_\nu}{4 \pi} 
\left[({\partial_x \phi_\nu})^2 - \Pi_\nu^2\right] \right\}
\  + \  g_{1 \perp} \ {\rm const.} \cos \sqrt{8 \pi} \phi_s \label{boson-ham},
\end{eqnarray}
where
$g_c = g_{2 \parallel} + g_{2 \perp}$ and 
$g_s = g_{2 \parallel} - g_{2 \perp}$.
The charge interaction $g_c$
can be absorbed into the free Hamiltonian by a
simple rescaling of the charge boson, but the  spin
interactions  $g_s$ and $g_{1 \perp}$ obey Kosterlitz-Thouless renormalization
group equations \cite{Kosterlitz} 
with flow lines along hyperbolas $g_s^2-g_{1 \perp}^2 = \rm const.$
(to lowest order).  For $g_s > -|g_{1 \perp}|$ the spin sector develops 
a gap in the
low energy, long wave-length limit, but for $g_s < -|g_{1 \perp}|$ 
the system flows
to a stable fixed point $g_s^* = -\sqrt{g_s^2 -g_{1 \perp}^2}, 
\ g_{1 \perp}^* = 0$.  For
$g_s = -|g_{1 \perp}|$ the interaction corresponds to one single marginally
irrelevant operator, so that $g_s^*=g_{1 \perp}^*=0$.
If the flow to a stable fixed point occurs, we can rescale the bosons 
by a canonical transformation to obtain a free theory ($\nu= s,c$)
\begin{equation}
\label{rescaledfields}
\phi_\nu \to {K_\nu \phi_\nu}, \ \ \ \Pi_\nu \to {\Pi_\nu / K_\nu},
\end{equation}
where to first order in the coupling constants
\begin{equation}
\label{Knu} 
K_s^2 = 1 - {g_s^*}/{2 \pi v_s}, \ \ \ K_c^2 =1-{g_c}/{2 \pi v_c}.
\end{equation}
This yields the Hamiltonian 
\begin{equation} \label{Ham}
{\cal H}  = \sum_{\nu=s,c}  \frac{\bar{v}_\nu}{2}
\left[({\partial_x \phi_\nu})^2 + \Pi_\nu^2\right] 
\end{equation}
where $\bar{v}_\nu=(v_\nu + \frac{g_\nu}{2 \pi}) K_\nu^2$, 
i.e.~$\bar{v}_\nu=v_\nu$ to first
order in the coupling constants so we will omit
the `bar' in the following. 

The chiral components of the electron field can now be expressed 
in terms of 
free boson fields and their duals by using Eq.~(\ref{JRLcsdef})
and the formula~\cite{Mandelstam}
\begin{equation}
\psi_{L/R,\sigma}(x) = \frac{\eta_\sigma}{\sqrt{2 \pi \alpha}} 
\exp{ \left( \mp 2 \pi i \int J_{L/R}^{\sigma}(x) dx \right)} \, ,
\label{formula}
\end{equation}
where $\eta_\sigma$ obeys the anticommutation relation 
$\{\eta_\uparrow,\eta_\downarrow\}=0$, with
$\eta_\sigma^2=1$. The presence of $\eta_\sigma$ in (\ref{formula}) 
guarantees that operators with different spin $\sigma$ obey anticommutation 
relations.

Using the duality relation 
$\Pi_\nu \equiv \partial_{{v}_\nu t} \phi_\nu=\partial_x \tilde{\phi}_\nu$, we 
thus obtain, using  Eqs. (\ref{JRLcsdef}) and (\ref{rescaledfields})
\begin{equation}
\label{eqn:ourbosform}
\psi_{L/R,\sigma}(x)   \propto \eta_\sigma  \prod_{\nu = c,s}
\exp{\left[  i \epsilon_{\nu,\sigma}
\sqrt{\frac{\pi}{2}} \left( \mp K_\nu \phi_\nu(x) -
 K_\nu^{-1} \tilde{\phi}_\nu (x) \right) \right]} \, , \label{bos.form}
\end{equation}
with $\epsilon_{\nu,\sigma}=1$ unless $\nu=s$ and $\sigma=\downarrow$ 
when it is equal to $-1$. Note that we have obtained the bosonization 
formula (\ref{eqn:ourbosform}) with {\em no} assumption about boundary 
conditions.

We now apply the formalism above to a system of length $L$ with  
{\it open} boundary conditions and thus require the electron 
field $\Psi_{\sigma}(x)$ to vanish at $x=0$ and at $x=L$. From 
Eq.~(\ref{psileftright}) we see that this implies  
\begin{mathletters} \label{fermionboundcond}
\begin{eqnarray}
\psi_{L,\sigma}(0,t) & = & - \psi_{R,\sigma}(0,t) \label{fermionboundcond0} \\
\psi_{L,\sigma}(L,t) & = & - e^{i 2 k_F L} \psi_{R,\sigma}(L,t) \, .
\label{fermionboundcondL}
\end{eqnarray}
\end{mathletters}
Considering Eq.~(\ref{bos.form}) this gives us fixed boundary conditions 
on the boson fields $\phi_\nu$, which in turn determine the mode expansion. 

To calculate the mode expansion for the boson we find it most convenient
to consider the classical Euler-Lagrange equation with fixed boundary
conditions at $x=0, L$
\begin{equation}
\phi_\nu (0,t)=C_0 \, , \ \ \
\phi_\nu (L,t)=C_L, \label{fixedBC}
\end{equation}
and then performing a canonical quantization. 
We therefore consider the  classical Lagrangian density corresponding
to the Hamiltonian (\ref{Ham})
\begin{equation} \label{Lagr}
{\cal L}  = \sum_{\nu=s,c}  \frac{{v}_\nu}{2} \left[ 
({\partial_{{v}_\nu t} \phi})^2 - ({\partial_x \phi})^2
\right] 
\end{equation}
where $\partial_{{v}_\nu t} \equiv \frac{1}{{v}_\nu} \partial_t$.
The resulting Euler-Lagrange equations can be expressed
in terms of $\partial_\pm =
\partial_x \pm \partial_{{v}_\nu t}$,
\begin{equation} \label{EulerLagr}
(\partial_{{v}_\nu t}^2 - \partial_x^2)\phi_\nu=
\partial_+ \partial_- \phi_\nu=0 \, ,
\end{equation}
and it follows that the two solutions 
can be written in terms of left- and right
moving bosons, $\phi_{\nu,L}(x+{v}_\nu t)$ and 
$\phi_{\nu,R}(x-{v}_\nu t)$.
We use the combination $\phi_\nu (x,t) = \phi_{\nu,L}(x+{v}_\nu t)+
\phi_{\nu,R}(x-{v}_\nu t)$ and its dual field 
$\tilde{\phi}_\nu (x,t)=
\phi_{\nu,L}(x+{v}_\nu t)-\phi_{\nu,R}(x-{v}_\nu t)$, 
related by 
\begin{equation} \label{duality}
\partial_{x} \phi_\nu  =  
\partial_{{v}_\nu t} \tilde{\phi}_\nu \, , \ \ \ \ \ 
\partial_{x} \tilde{\phi}_\nu  =  
\partial_{{v}_\nu t} \phi_\nu \, .
\end{equation}

The classical solution with the boundary condition (\ref{fixedBC}) is
obtained in a straightforward way, and after canonically quantizing
we find the mode expansion
for the quantum fields according to the boundary condition 
(\ref{fermionboundcond})
\begin{mathletters}
\begin{eqnarray}
\phi_{\nu} (x,t) & = & \phi_{\nu,0} + \hat{Q}_\nu \frac{x}{ L} 
+ \sum_{n=1}^\infty \frac{1}{\sqrt{n \pi}}\sin{\frac{n \pi x}{L}} 
\left( - i e^{- i \frac{n \pi  v_{\nu}t}{L} } a_n^{\nu} 
+ \mbox{h.c.} \right) \, , \\
\tilde{\phi}_{\nu} (x,t) & = & \tilde{\phi}_{\nu,0} +
\hat{Q}_\nu \frac{v_{\nu}t}{ L} 
+ \sum_{n=1}^\infty \frac{1}{\sqrt{n \pi}} \cos{\frac{n \pi x}{L}}
\left( e^{- i \frac{n \pi v_{\nu}t}{L}} a_n^{\nu} 
+ \mbox{h.c.} \right) \, .
\end{eqnarray}
\label{EXPANSION}
\end{mathletters}
The non-zero commutation relations among the mode operators are 
$[a_n^{\nu},{a_n^{\nu}}^{\dagger}]=1$ and 
$[\tilde{\phi}_{\nu,0},\hat{Q}_\nu]=i$, 
while $\phi_{\nu,0}$ are c-numbers. 
This result (\ref{EXPANSION}) obtained from the classical
solution {\it automatically} contains the correct 
zero modes, in particular the total charge and spin operators $\hat Q_\nu$,
which were first postulated by Haldane for a periodic system\cite{Haldane}.
The resulting energy spectrum from Eq.~(\ref{Ham}) is
\begin{equation} \label{spectrum}
H = \sum_{\nu=s,c} \left( \frac{v_\nu}{2 L}\hat{Q}_\nu^2 +
\sum_{n=1}^\infty \frac{\pi v_\nu n}{L}
{a_{n}^{\nu}}^\dagger a_{n}^{\nu}  \right) \ .
\end{equation}

From Eqs.~(\ref{EXPANSION}) we can read off 
the mode expansions for left-moving bosons
\begin{eqnarray}
\label{eqn:Lmodeexp}
\phi_{\nu,L} (x,t)& \equiv &\frac{1}{2}\left[ \phi_{\nu} (x,t) +
\tilde{\phi}_{\nu} (x,t)\right] \nonumber \\
&=& \frac{1}{2}\left( \phi_{\nu,0}+\tilde{\phi}_{\nu,0} \right)
+\hat{Q}_\nu \frac{x+v_{\nu}t}{2L}+\sum_{n=1}^\infty \frac{1}{\sqrt{n \pi}}
\left[ e^{- i \frac{n \pi \left(x+v_{\nu}t\right)}{L}} a_n^{\nu}
+ \mbox{h.c.} \right] \, .
\end{eqnarray}
The right-moving boson field can be related to the left-moving one by
\begin{equation}
\phi_{\nu,R} (x,t) \equiv \frac{1}{2}\left[ \phi_{\nu} (x,t)-
\tilde{\phi}_{\nu} (x,t)\right]=
- \phi_{\nu,L} (-x,t) + \phi_{\nu,0} . \label{phiR}
\end{equation}

The boundary conditions (\ref{fermionboundcond}) on the fermions
provide us with the quantization condition for the eigenvalues of the 
operators  $\hat Q_\nu$
(using special commutation relations
of $\phi_{\nu}$ and $\tilde{\phi}_{\nu}$ at the boundary\cite{WongAffleck}
that follow from the mode expansion (\ref{EXPANSION}))
\begin{mathletters}
\begin{eqnarray}
K_c Q_{c} & = & \sqrt{\frac{\pi}{2}}
\left(n +1 + \frac{2 k_F L}{\pi} \right) \ , \\
K_s Q_{s} & = & \sqrt{\frac{\pi}{2}} m \ , 
\end{eqnarray}
\label{nontrivial}
\end{mathletters}
where $m$ and $n$ are either both odd or both even integers. Therefore,
the quantum numbers for the total spin $m$ and the total
charge $n$ are {\it not} independent, which is to be expected
because we can only insert and remove real electrons, i.e. we cannot 
change the total charge and the
total magnetization independently. In this sense these degrees of freedom
do {\it not} obey spin-charge separation and the partition function does 
not factorize.  It is interesting to note the formal similarity to a {\it 
spinless periodic} system, which is also described by two channels (left
and right moving) and
where a similar condition holds for the total current and charge 
quantum numbers\cite{Haldane}. As we will see later, the zero modes
indeed give a contribution to the Green's function which does not
factorize, while all dynamical degrees of freedom
in Eq.~(\ref{spectrum}) remain spin-charge separated.

The constants  $\phi_{c,0}=\sqrt{\frac{\pi}{2}} K_c^{-1}$ and $\phi_{s,0}=0$ 
are also determined by the boundary condition (modulo the intrinsic 
periodicity of the boson $\sqrt{2 \pi} K_\nu^{-1}$). Hence,
 we obtain from Eqs. (\ref {bos.form}) and (\ref{phiR})
\begin{equation} \label{LR-relate}
\psi_{R,\sigma} (x,t) = - \psi_{L,\sigma} (-x,t) \, .
\end{equation}
which allows us to write the full theory in terms of left-movers only. 
This concludes our analysis of bosonization in the presence of open boundaries.
For an alternative approach - exploiting the path integral formulation - 
see Ref. \onlinecite{fradkin}.
 
\section{Green's functions}
Using the formalism above, the exact single electron 
Green's function for a confined Luttinger liquid with open boundaries
can now be calculated. With the decomposition in 
Eq.~(\ref{decomposition}) and using Eq.~(\ref{LR-relate}), we have
\begin{eqnarray}
 \left<\Psi^{\dagger}_\sigma(x,t) \Psi^{}_\sigma(y,0) \right>  
=& &  e^{i k_F (x-y)}G(x,y,t) \ + \ e^{-i k_F (x-y)}G(-x,-y,t) \nonumber \\
&  -&   e^{i k_F (x+y)}G(x,-y,t) \ - \
e^{-i k_F (x+y)}G(-x,y,t),  \label{fullGF}
\end{eqnarray}
where the chiral Green's function 
$G(x,y,t) \equiv \left<\psi^{\dagger}_{L,\sigma}(x,t) \psi^{}_{L,\sigma}(y,0) 
\right>$
is derived in Appendix A. The result is a product of the spin and charge
contributions $F_{s,c}$ and a factor $H$  from the zero modes 
\begin{eqnarray}
G(x,y,t) \ \propto H(x,y,t)
\prod_{\nu = c,s} & &
\left(F_\nu(v_\nu t+ x-y) \right)^{-\frac{(K_\nu+K_\nu^{-1})^2}{8}}
\left(F_\nu(v_\nu t- x+y) \right)^{-\frac{(K_\nu-K_\nu^{-1})^2}{8}}
 \label{finiteLfiniteTGF} \\ & \times &
\left( \frac{|F_\nu (2 x)|| F_\nu (2y)|}
{F_\nu (v_\nu t+x+y) F_\nu (v_\nu t-x-y)}
\right)^\frac{K_\nu^{-2}-K_\nu^{2}}{8} \, . \nonumber
\end{eqnarray}
The contribution $H$ from the zero modes is given by (see Appendix A)
\begin{eqnarray} \label{H}
H(x,y,t)&=&
\frac{\vartheta_2(u_c+ \tau_c k_F L| \tau_c)
\vartheta_3(u_s|\tau_s)+\vartheta_3(u_c+ \tau_c k_F L| \tau_c)
\vartheta_2(u_s|\tau_s) }{\vartheta_2( \tau_c k_F L| \tau_c)
\vartheta_3(0|\tau_s)+\vartheta_3( \tau_c k_F L| \tau_c)
\vartheta_2(0|\tau_s)}e^{i 2 u_c \frac{k_F L}{\pi}}
\end{eqnarray}
where the Theta functions, $\vartheta_k(u|\tau)$, are defined in 
sections~8.18-8.19 in~\cite{GR} and
\begin{equation}
u_\nu =-\frac{\pi}{2}\left(\frac{v_\nu  K_\nu^{-2}t + x-y}{L}\right)\ \ \ \ \ 
\tau_\nu = i \frac{v_\nu \beta}{K_\nu^2 L}.
\label{utau}
\end{equation}
The expression for $H$ does not contain any poles, but 
may still influence the spectral properties significantly
in mesoscopic systems\cite{loss,EMK}.  
It is also interestingly to note that the contribution from the zero modes
{\it cannot} be written as a product of independent spin and charge
contributions, which is expected since the quantum numbers $n$ and
$m$ in Eq.~(\ref{nontrivial}) are not independent.

The factors $F_\nu$ from the spin and charge bosonic modes can also be 
written in terms of the Theta functions (see sections~8.18-8.19 in~\cite{GR})
\begin{equation} \label{boseexpval}
F_\nu(z)=i {\frac{2L}{\alpha \pi} \sin\frac{\pi z}{2L}}
\prod_{k=1}^\infty \left[1+\left(\frac{\sin\frac{\pi z}{2L}}
{\sinh{k\frac{\pi v_\nu \beta}{2L}}}\right)^2 \right] \ = \ 
\frac{\vartheta_1(\frac{\pi z}{2 L}|i\frac{v_\nu \beta}{2 L})}{\vartheta_1(
-i\frac{\pi \alpha}{2 L}|i\frac{v_\nu \beta}{2 L})}\, .
\end{equation}
Here $x$ and $y$ denote the distance from the left boundary ($x=0$) and the
argument $z$ carries an implicit cut-off $z - i \alpha$. The parameters 
$v_\nu$ and $K_\nu$ are defined in (\ref{vnu}) and (\ref{Knu}) respectively. 

We can immediately verify that the full Green's function is 
anti-periodic under translation by the inverse temperature 
$t \to t+i\beta$ as it should be.  It is also
interesting to note that the factor from the zero modes is
periodic under the change of the Fermi-level
$k_F$ by $\pi/L$, which results in periodic oscillations of the spectral
properties of the system as the chemical potential is changed. This
is a manifestation of Coulomb-blockade oscillations, i.e.~resonances 
can be observed if the system is of ``mesoscopic'' size\cite{EMK}.
It is important to emphasize that the zero mode contribution $H$
in Eq.~(\ref{H}) is obtained by using a {\it grand} canonical 
ensemble when taking the averages, thus allowing for fluctuations
in the magnetization $m$ and in the particle number $n$.
By fixing these quantum numbers (i.e. using an idealized ``closed''
system) {\em or} by letting the system size tend to infinity, the 
zero mode contribution collapses to a constant phase.
This provides a vivid example of how different statistical ensembles may lead
to different results on mesoscopic scales and downwards, where
quantum coherence effects become important\cite{Kamenev}.

We see in Eq.~(\ref{finiteLfiniteTGF}) that we recover the universal 
power-laws, which give the expected branch cuts in the Green's 
function\cite{Voit}.
In addition, we get a contribution from the boundary, which is entirely
contained in the last factor of Eq.~(\ref{finiteLfiniteTGF}) and gives an
additional analytic structure.  This factor does not contribute
in non-interacting systems ($K_c = K_s = 1$), so that in this case
the presence of the boundary 
is seen only by the addition of the two last ``Friedel'' terms in 
Eq.~(\ref{fullGF}) compared to the bulk case.
In contrast, {\em with} interaction ($K_s,\, K_c < 1$), the boundary 
influences also the chiral Green's function. 
This is expected, since only {\em with} interactions can effects from 
electron scattering off the boundary propagate to other parts of the 
system, thus influencing also the chiral pieces of the full Green's function.   

The zero-temperature limit $T \rightarrow 0$ is readily obtained by letting 
$\beta \rightarrow \infty$ in Eqs.~~(\ref{H}) and (\ref{boseexpval})
\begin{eqnarray}
G(x,y,t) \ \propto \ e^{i(2 n_0-1) u_c}
\prod_{\nu = c,s} &&
\left({\frac{2L}{\pi}\sin\frac{\pi(v_\nu t+ x-y)}{2L}}
\right)^{-\frac{(K_\nu+K_\nu^{-1})^2}{8}}
\left({\frac{2L}{\pi}\sin\frac{\pi(v_\nu t- x+y)}{2L}}
\right)^{-\frac{(K_\nu-K_\nu^{-1})^2}{8}}\nonumber
\\ & \times &
\left( \frac{\sin\frac{\pi x}{L}\sin\frac{\pi y}{L} }
{\sin\frac{\pi(x+y+v_\nu t)}{2L}
\sin\frac{\pi(x+y-v_\nu t)}{2L}}
\right)^\frac{K_\nu^{-2}-K_\nu^{2}}{8}
\label{finiteLGF} 
\end{eqnarray}
where $n_0 = (k_FL/\pi\ {\rm mod}\ 1)$ effectively measures the difference
between the Fermi-vector and the highest occupied level (which are not
necessarily the same in a system with discrete energy levels).
The phase $e^{i(2 n_0-1) u_c}$ comes from the zero modes and {\it does}
influence the time correlations.  For special values of $n_0$ this
phase may have a different dependence on $u_\nu$.  When this phase
is neglected, Eq.~(\ref{finiteLGF}) agrees  with the results in 
Ref.~\onlinecite{fabrizio}, obtained via  bosonization according to the 
``Haldane prescription''\cite{Haldane}.  It is also in 
agreement with results~\cite{OurPRL}, obtained by conformally 
mapping the semi-infinite complex plane onto a finite strip. 

Using the Poisson summation formula, we can obtain the limits 
of the Theta functions
$\vartheta_1(z|i \gamma) \approx 2 \gamma^{-1/2} e^{-z^2/\pi \gamma}
e^{-\pi/4 \gamma} \sinh \frac{z}{\gamma}$ and
$\vartheta_2(z|i\gamma) \approx \vartheta_3(z|i\gamma) \approx
\gamma^{-1/2} e^{-z^2/\pi \gamma}$ as $\gamma \to 0$. Hence, 
by letting $L \rightarrow \infty$ in Eqs.~(\ref{H}) and 
(\ref{boseexpval}),
we obtain the finite temperature chiral
Green's function of a semi-infinite system with open boundary
\begin{eqnarray}
G(x,y,t) \ \propto 
\prod_{\nu = c,s} & & 
\left({\frac{v_\nu \beta}{\pi}\sinh\frac{\pi(v_\nu t+ x-y)}{v_\nu \beta}}
\right)^{-\frac{(K_\nu+K_\nu^{-1})^2}{8}}
\left({\frac{v_\nu \beta}{\pi}\sinh\frac{\pi(v_\nu t- x+y)}{v_\nu \beta}}
\right)^{-\frac{(K_\nu-K_\nu^{-1})^2}{8}}\nonumber \\  & \times &
\left( \frac{\sinh\frac{\pi 2 x}{v_\nu \beta}\sinh\frac{\pi 2 y}{v_\nu \beta}}
{\sinh\frac{\pi(x+y+v_\nu t)}{v_\nu \beta}
\sinh\frac{\pi(x+y-v_\nu t)}{v_\nu \beta}}
\right)^{\frac{K_\nu^{-2}-K_\nu^{2}}{8}} \ .  
 \label{finiteTGF} 
\end{eqnarray}

Finally, the $T \rightarrow 0$ limit for a semi-infinite system is obtained by
letting $\beta \rightarrow \infty$ in Eq.~(\ref{finiteTGF})
({\em or} $L \rightarrow \infty$ in Eq.~(\ref{finiteLGF})):
\begin{eqnarray}
G(x,y,t) \ \propto 
\prod_{\nu = c,s} & &
\left({v_\nu t+ x-y}\right)^{-\frac{(K_\nu+K_\nu^{-1})^2}{8}}
\left({v_\nu t- x+y}\right)^{-\frac{(K_\nu-K_\nu^{-1})^2}{8}}
\left( \frac{4 x y} {[(x+y)^2-v_\nu^2 t^2]}
\right)^\frac{K_\nu^{-2}-K_\nu^{2}}{8} \ . 
\label{T0Linfty} 
\end{eqnarray}
We note that in the limit $xy \gg |(x-y)^2 - v_\nu^2 t^2|$
the last factor in equation (\ref{T0Linfty})
goes to unity and we recover the known zero-temperature bulk correlation function\cite{Voit}
(as we also do in the non-interacting case $K_c = K_s = 1$). In the limit of 
equal time, $t=0$, this gives the asymptotic correlator
\begin{equation}
G(x,y,t=0) \propto \frac{1}{(x-y)^{2\Delta}}
\label{bulk}
\end{equation}
with the equal-time bulk exponent $\Delta =  (K^2_c + K^2_s + K^{-2}_c + 
K^{-2}_s)/8$. In contrast, when one of the points is close to the 
boundary compared to the relative distance $x \gg y$ one finds 
\begin{equation}
G(x,y,t=0) \propto \frac{1}{(x-y)^{2\Delta_{\perp}}} , \ \ \ \ 
\label{perpendicular}
\end{equation}
with $\Delta_{\perp} =  (3K^{-2}_c + 3K^{-2}_s +K^{2}_c + 
K^{2}_s)/16$. This is to be compared to the asymptotic long-time behavior of 
the autocorrelation function in the limit $t \gg x,y$, which behaves as
\begin{equation}
G(x,x,t) \propto \frac{1}{t^{2\Delta_{\parallel}}}
\label{parallel}
\end{equation}
with $\Delta_{\parallel} =  (K^{-2}_c + K^{-2}_s)/4$. As expected from scale 
invariance, we thus recover the scaling law $\Delta_{\perp} =  
(\Delta + \Delta_{\parallel})/2$. It may be worth pointing out that the {\em 
dynamic} ($t \neq 0$)
 asymptotic large distance correlator in Eq.~(\ref{perpendicular})
is {\em not} governed by the single 
exponent $\Delta_{\perp}$, as one may naively have expected from the 
analogy with classical critical phenomena \cite{Diehl}. Instead it 
remains a product of separate charge- and spin 
correlators with exponents $ (3K^{-2}_c + K^{2}_c)/16$ and 
$ (3K^{-2}_s + K^{2}_s)/16$, respectively. 
This is not in conflict with scale invariance since the theory is built out 
of two distinct sectors, each with its own effective velocity. More 
importantly, the behavior 
in Eq.~(\ref{parallel}) reveals that {\em the asymptotic 
low-energy behavior of a Luttinger liquid with an open boundary belongs 
to a different universality class than that of the bulk theory.} We shall 
elaborate on this in the next section when we discuss the local density 
of states.

\section{Local Spectral Density}

To understand the physical implications of the boundary correlations
we study the {\it local} spectral density $N(\omega,r)$,
given in terms of the single electron Green's function in (\ref{fullGF})
\begin{equation}
N(\omega,r)  =  \frac{1}{2 \pi} \int_{-\infty}^\infty
\!\!e^{i \omega t} \left<\left\{
\Psi^{\dagger}_\sigma(r,0), \Psi^{}_\sigma(r,t)\right\}\right> dt,
\label{N_omega}
\end{equation}
where $\omega$ is measured relative to the Fermi energy and
$r$ is the distance from the boundary.

At $T=0$ and without the boundary, the integral in (\ref{N_omega}) can be done 
exactly\cite{MedenSchonhammerVoit}
and one finds that the spectral density scales at the
Fermi level as $N(\omega) \propto \omega^{\alpha_{\rm bulk}}$,
where the exponent in the bulk is given by 
\begin{equation}
\alpha_{\rm bulk} = (K_c^2 + K_c^{-2} + K_s^2 + K_s^{-2})/4 -1. \label{alpha}
\end{equation}

However, the boundary clearly influences this scaling behavior,
and by inspection of Eq.~(\ref{T0Linfty}) for a semi-infinite system at $T=0$, 
simple power counting reveals that there must be a  crossover 
to a boundary dominated regime for $r \omega < v_c, v_s$ with a novel
exponent 
\begin{equation}
\alpha_{\rm bound} = (K_c^{-2} + K_s^{-2})/2 -1\ . 
\end{equation}
Interestingly, the boundary exponent $\alpha_{\rm bound}$ therefore {\it always}
dominates for sufficiently small $\omega$.  It is also interesting to note that
the last two terms in equation (\ref{fullGF}) make a contribution which
oscillates at twice the Fermi wave-vector and drops off with
the distance from the boundary proportional to
$e^{i 2 k_F r} r^{-(K_c^{-2} +K_s^{-2})/2}$.  This contribution
is reminiscent of a Friedel oscillation, although it can 
probably not be observed directly, since
experimental measurements of the density of states 
(in particular photoemission)
effectively averages over several lattice sites.  We therefore ignore those 
``Friedel'' terms in the following calculations and make the replacement 
\begin{equation}
\left< \left\{
\Psi^{\dagger}_\sigma(r,0), \Psi^{}_\sigma(r,t)\right\}\right> \rightarrow
G(r,r,-t)+G(-r,-r,-t)+G(r,r,t)+G(-r,-r,t) \, ,
\label{rep}
\end{equation}
in Eq.~(\ref{N_omega}). Using our exact results for $G(x,y,t)$ in the previous 
section allows us to fully explore the physically relevant piece of the 
local spectral density.  

\subsection{The zero temperature and infinite length limit}   
\label{T0L0}

For $T \rightarrow 0$ and a semi-infinite system, $L \rightarrow \infty$, 
it is readily derived from (\ref{T0Linfty}), (\ref{N_omega}) and 
(\ref{rep}) that
\begin{eqnarray} \label{NT0Linf}
&N&(\omega,T=0,r)\ =\ 
\frac{2}{\alpha \pi^2} v_c^{-a_c} v_s^{-a_s} 
\int_0^\infty{dt} \cos{\gamma(t)} \ 
\left[\cos{\omega t}-1\phantom{\frac{}{}}\right] \ 
\left(\frac{t}{\alpha}\right)^{-a_s-a_c}
\left|1-\left(\frac{v_c t}{2 r}\right)^2\right|^{-b_c/2} 
\left|1-\left(\frac{v_s t}{2 r}\right)^2\right|^{-b_s/2} 
\end{eqnarray}
where
\begin{equation} \label{gamma}
\gamma(t)= \left\{ \begin{array}{lcl}
               \frac{\pi}{2} (a_s+a_c) &\phantom{nnn}& 0<t<\frac{2r}{v_c}\\
               \frac{\pi}{2}(a_s+a_c+b_c) && \frac{2r}{v_c} < t < \frac{2r}{v_s} \\
               \frac{\pi}{2} (a_s+a_c+b_s+b_c) && \frac{2r}{v_s} < t < \infty
                  \end{array} 
           \right.
\end{equation}
and
\begin{equation}
a_\nu = \frac{K_\nu^2 + K_\nu^{-2}}{4}, \ \ \ \  
b_\nu = \frac{K_\nu^{-2} - K_\nu^2}{4}. \ \ \ \ 
\label{kabdef}
\end{equation}
To cure the original divergences in (\ref{NT0Linf}) we have 
subtracted an infinite 
constant by renormalizing the static spectral density to zero for any given 
$r$, i.e.~$N(\omega=0, T=0, r) = 0$. 

Let us study the two limiting cases $r \rightarrow \alpha$ ({\em 
boundary regime}, with $\alpha$ the short-distance cutoff) and 
$r \rightarrow \infty$ 
({\em bulk regime}). Both limits are easily obtained from Eq.~(\ref{NT0Linf}) 
by using the integral (3.823 in \cite{GR})    
\begin{equation}
\int_0^\infty dx (\cos{x} -1)x^{-k} = \frac{\pi}{\Gamma(k)} 
\frac{1}{2 \cos{\frac{\pi}{2}k}}
\ \ \  \ 1<k<3 .
\end{equation}
with $\Gamma(k)$ the Gamma function. This gives
\begin{equation} \label{Nextreme}
N(\omega)=\left\{ \begin{array}{lcl}
\frac{\alpha^{\alpha_{\rm bulk}} v_c^{-a_c} v_s^{-a_s} }{\pi
\Gamma(1+\alpha_{\rm bulk})} \ 
\omega^{\alpha_{\rm bulk}} &
\phantom{nnn} & r \to \infty \\
\frac{\alpha^{\alpha_{\rm bound}} v_c^{-a_c-b_c} v_s^{-a_s-b_s} }{\pi
\Gamma(1+\alpha_{\rm bound})} \  
\omega^{\alpha_{\rm bound}} &
\phantom{nnn} & r \to \alpha \end{array} \right.
\end{equation}
where $\alpha_{\rm bulk}= a_s+a_c-1$
and  $\alpha_{\rm bound}= a_s+a_c+b_s+b_c-1$. 
In the bulk limit $r \to \infty$ this result 
is in full agreement with previous 
calculations\cite{MedenSchonhammerVoit,Suzumura}, but in the boundary 
limit $r \to \alpha$ we observe completely different exponents.
By substituting $\omega t$ by $x$ in the integral in Eq.~(\ref{NT0Linf})
we see that apart from a prefactor measuring the distance to the boundary, 
the spectral density depends only on the scaling variable 
${r \omega}$. This implies that the condition for boundary behavior is
$\frac{r \omega}{v_c} \ll 1$, 
{\em i.e.~regardless of the value of $r$ there will always be a region
in $\omega$ around the Fermi-energy 
where the spectral density is determined by the boundary
exponent $\alpha_{\rm bound}$}. 
Actually, the scaling behavior of the spectral density splits into
three distinct regions in $\omega$ where different exponents govern the 
{\em leading} scaling. 

\begin{figure}
\begin{center}
\mbox{\epsfxsize=4.5in \epsfbox{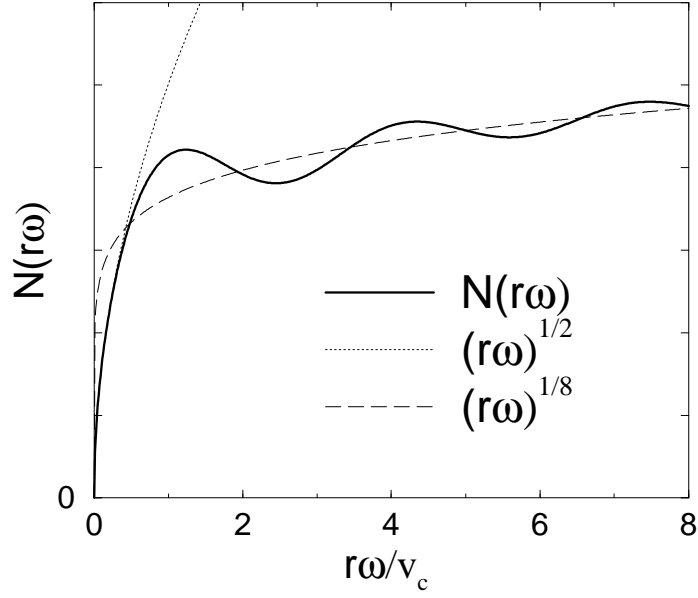}}
\end{center}
\caption{The spectral density as a function of $r\omega$ in arbitrary
units (from Ref. \protect{\onlinecite{OurPRL}}). The corresponding
power-laws for $\alpha_{\rm bulk} = 1/8$ and $\alpha_{\rm bound}= 1/2$ 
are also shown.
The distance from the boundary $r$ is held constant and just fixes the scale.}
\label{n_ro}
\end{figure}

As an example, let us choose parameters adapted to a description of 
the large-$U$ Hubbard chain away from half-filling 
(cf. Sec.~II and Ref.~\onlinecite{Voit}). 
In this case the SU(2)-invariance forces $K_s =1$, and
it is known from Bethe ansatz calculations that $K_c^2 \to 1/2$ when
$U \to \infty$\cite{Schulz,LiebWu}. 
Therefore, the spin channel is not affected by the boundary (since $b_s=0$), 
and the local 
spectral density $N(\omega,r)$ in Eq.~(\ref{NT0Linf}) splits into two
asymptotic sectors only: 
The boundary regime for $\omega \ll \frac{v_c}{r}$ and the
bulk regime for $\omega \gg \frac{v_c}{r}$. From these numbers, 
the well known result $N(\omega) \propto |\omega|^{1/8}$
follows immediately for the bulk regime, as can be seen 
from Eq.~(\ref{Nextreme}).   
In the presence of the boundary, however, we cross over to the boundary
exponent $\alpha_{\rm bound} = 1/2$ for $\omega < v_c/r$.  
The results
in Fig.~(\ref{n_ro}) clearly show the crossover from boundary behavior 
for $r \omega/v_c < 1$ with exponent $\alpha_{\rm bound}=1/2$ 
to bulk behavior for $r \omega/v_c > 1$ with
exponent $\alpha_{\rm bulk}=1/8$ (the corresponding
power-laws are superimposed). In the figure, the distance from the 
boundary, $r$, is held constant, thus setting
the scale. The observed oscillations in Fig.~(\ref{n_ro}) 
are an intriguing secondary effect, which
vanish asymptotically as $\sin (2 \omega r/v_c) (\omega r)^{b_c/2-1}$. It 
is important to emphasize that they are {\it not} due to the ``Friedel'' 
terms in equation (\ref{fullGF}) which have been neglected. Instead, 
they originate from the integrable singularity
of the integrand in Eq.~(\ref{NT0Linf}) at $t={2r \over v_c}$, 
which is only present in the boundary case.

\subsection{The finite temperature and infinite length limit}
\label{TL0}

We now consider the effect of finite temperatures on a semi-infinite system 
with a boundary. In general, by turning on temperature one induces 
a different behavior for small $\omega$: Both
in bulk and boundary regimes (defined as above) the spectral density 
crosses over to $\omega^2$ scaling when $\omega < 2\pi/\beta $. 
This is due to thermal fluctuations which produce exponential damping of 
the density correlations for unequal times. 
The formal expression for $N(\omega,\beta,r)$ 
for this case, with $\beta = 1/k_BT$ and $k_B=1$ is given by
\begin{eqnarray} \label{NTLinf}
 &N&(\omega, \beta,  r)\ =\ \frac{2}{\alpha \pi^2} 
 v_c^{-a_c} v_s^{-a_s} 
\int_0^{\infty}dt \cos{\gamma(t)}\left[ \cos{\omega t}\ 
\left(\frac{\sinh{\frac{\pi}{\beta}t}}{\frac{\pi}{\beta}\alpha}
\right)^{-a_s-a_c}
\left|\frac{\sinh{\frac{\pi}{v_c\beta}(2r+ v_c t)}
\sinh{\frac{\pi}{v_c \beta}(2r- v_c t)}}{\sinh^2{\frac{2 \pi r}{v_c \beta}}}
\right|^{-b_c/2}\right.
\nonumber \\& &\times 
\left|\frac{\sinh{\frac{\pi}{v_s \beta}(2r+v_s t)}
\sinh{\frac{\pi}{v_s \beta}(2r-v_s t)}}{\sinh^2{\frac{2 \pi r}{v_s \beta}}}
\right|^{-b_s/2}
\left. - 
\left(\frac{t}{\alpha}\right)^{-a_s-a_c}
\left|1-\left(\frac{v_c t}{2 r}\right)^2\right|^{-b_c/2} 
\left|1-\left(\frac{v_s t}{2 r}\right)^2\right|^{-b_s/2}\right],
\end{eqnarray}
where $\gamma(t)$ is given in Eq.~(\ref{gamma}).
We have subtracted the same infinite constant as in Eq.~(\ref{NT0Linf}), 
thereby, as before, renormalizing the zero-temperature 
static spectral density to zero for any given 
$r$, i.e.~$N(\omega=0, T=0, r) = 0$.

\begin{figure}
\begin{center}
\mbox{\epsfxsize=4.5in \epsfbox{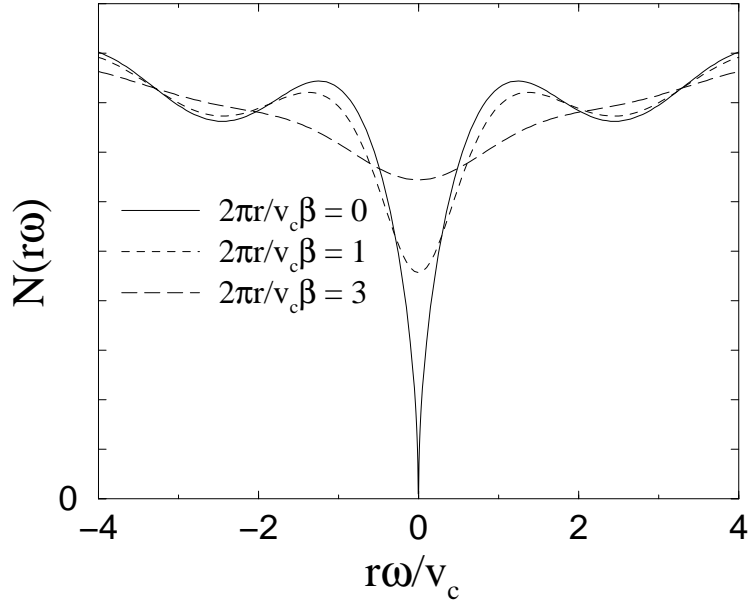}}
\end{center}
\caption{The spectral density as a function of $r\omega$ in arbitrary units
for different temperatures $r/\beta$.
For $\omega \ll 1/\beta$ we have parabolic behavior
which crosses over to the $T=0$ behavior for $\omega \gg 1/\beta$. }
\label{n_ro_roverbeta}
\end{figure}

Substituting $\omega t$ by $x$ in the integration of Eq.~(\ref{NTLinf}) 
as above reveals that 
$N(\omega,\beta,r)$ can be expressed as a function of 
{\it two} scaling variables ${r \omega}$ 
and ${\omega \beta}$ (up to the same $r-$dependent prefactor as 
for zero temperature). By inspection we recover the $T=0$ result when
$\frac{\omega \beta}{2 \pi} \gg 1$, but for small $\omega$ a new behavior 
sets in. Independently of whether we are in the boundary or bulk region, 
$N(\omega,\beta,r)-N(0,\beta,r)$ will be proportional to
$\omega^2$ for small $\omega \beta$
\begin{eqnarray}
N(\omega,\beta,r) - N(0,\beta,r) & \propto & 
\int_{0}^\infty dt(\cos{\omega t} -1) 
\left(\sinh{\frac{\pi}{\beta}t}\right)^{-k} \nonumber \\
& =& -2 \frac{\beta}{\pi}
\int_{0}^\infty dx \left( \sinh{x}\right)^{-k}
\sin^2{\frac{\omega \beta}{2 \pi}x} 
 \ \stackrel{\mbox{\tiny $\omega \beta$ small}}{\longrightarrow}  \
- \omega^2 \frac{1}{2} \left(\frac{\beta}{\pi}\right)^3 
\int_{0}^\infty dx  \left( \sinh{x}\right)^{-k} x^2
\end{eqnarray}
where the last integral converges if $0<k<3$ (in our case $k=\alpha_n+1$
where $\alpha_n$ is the boundary or bulk exponent). In conclusion,
the spectral properties are unaffected 
for energies well above the temperature $\omega \gg \frac{2 \pi}{\beta}$,
as expected.
However, the spectral density will exhibit a  $\omega^2$ behavior for 
$\omega < \frac{2 \pi}{\beta}$ before the
cross-over to the $T=0$ behavior occurs, as can be seen in 
Fig.~(\ref{n_ro_roverbeta}) where we again have considered 
the large-U Hubbard model away from half-filling. This is in complete
agreement with the recent work by Nakamura and Suzumura\cite{nakamura}
where the analogous effect was reported for an infinite ``bulk'' system.
Effectively, Fig.~(\ref{n_ro_roverbeta}) contains all information about both 
the bulk and the boundary
case, since we are free to adjust the distance from the boundary $r$ to
any value and this only changes the scale on which we measure the
energies and temperatures.  Therefore, we observe a cross-over from the 
quadratic behavior directly to bulk behavior if $\frac{\omega \beta}{2 \pi}
> 1$, while an intermediate boundary region can be observed for
$\frac{\omega \beta}{2 \pi} \alt 1$.
As we can see in  Fig.~(\ref{n_ro_roverbeta}), 
the spectral density can look very flat around the 
Fermi-level in either case,
and  the sharp cusp which has been predicted for T=0 
may not at all be visible in experiments.

\begin{figure}
\begin{center}
\mbox{\epsfxsize=4.3in \epsfbox{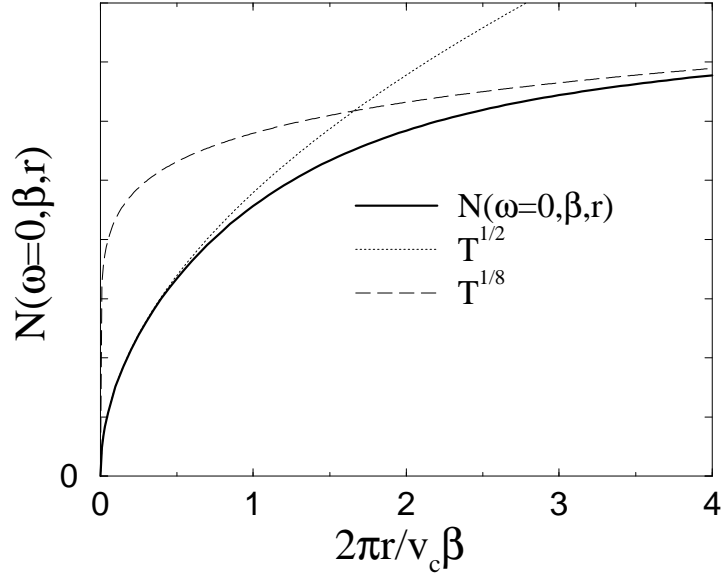}}
\end{center}
\caption{The spectral density at $\omega=0$ as a function of
$\frac{2 \pi r}{v_c \beta}$ in arbitrary units. The corresponding
power-laws for $\alpha_{\rm bulk} = 1/8$ 
and $\alpha_{\rm bound}= 1/2$ are also shown.
The distance $r$ to the boundary is held fixed and sets the scales.}
\label{n0_roverbeta}
\end{figure}

It is interesting to note that the boundary exponent also shows up in the
temperature dependence of the static spectral density 
$N(\omega=0,\beta,r)$. This is expected since the static density samples 
{\em all} times, with the asymptotic large-time behavior governed by the 
boundary exponent. As can be obtained from (\ref{NTLinf}), 
\begin{eqnarray}
N(\omega=0,\beta,r) &  = & 
\frac{2}{\pi^2} (\pi \alpha)^{a_s+a_c-1} v_c^{-a_c} v_s^{-a_s}
\nonumber \\
& & \hspace{1.5cm}\times \left\{ \begin{array}{lcl}
\beta^{-\alpha_{\rm bound}} 
(\frac{2 \pi r}{v_c})^{b_c} (\frac{2\pi r}{v_s})^{b_s} 
C(\alpha_{\rm bound}+1)& \phantom{nnn} &\frac{2 \pi r}{v_s \beta} \ll 1 \\
 \beta^{-\alpha_{\rm bulk}} C(\alpha_{\rm bulk}+1)& &
\frac{2 \pi r}{v_c \beta} \gg 1 
\end{array} \right.
\end{eqnarray}
where 
\begin{equation}
C(k)=\cos{\frac{\pi}{2}k} \int_0^{\infty} dx 
\left( \sinh^{-k}{x} - x^{-k} \right)
\end{equation}
which is convergent for $1<k<3$. With parameters again chosen to describe the 
large-$U$ Hubbard chain (away from half-filling), the boundary dominated 
regime opens up for $0 < \frac{2 \pi r}{v_c \beta} \alt 1$, as depicted in 
Fig.~(\ref{n0_roverbeta}).

\subsection{The zero temperature and finite length limit}

We now turn to a confined system with open boundaries at both ends.
At this point we want to emphasize that there is a
 distinction between effects that
arise from a non-trivial boundary condition (as discussed in Sec.~\ref{T0L0})
and effects from a finite system size 
(which may or may not have trivial boundary conditions).

A confined system with open boundary conditions
is technically more difficult to analyze, since the function  
$F_\nu(z)$ in Eq.~(\ref{boseexpval}) is periodic in $z$ with twice
the system size $2L$. Because of the non-integer exponents 
in Eq.~(\ref{finiteLGF}), we need to carefully keep track of the overall phase
as we integrate around the various  branch-cuts.  However, we
can simplify things by explicitly using the short distance
cutoff $z \rightarrow z - i\alpha$, which allows us to make a
Taylor expansion of the factors in Eq.~(\ref{finiteLGF}),
\begin{equation}\label{Taylorexp}
\left( i \sin{\frac{\pi z}{2 L}} \right)^\gamma=\left( \frac{e^{i 
\frac{\pi z}{2 L}}}{2} \right)^\gamma \sum_{n=0}^{\infty} c_n(\gamma) e^{-in 
\frac{\pi z}{L}}
\end{equation}
with
\begin{equation} \label{Taylorcoeff}
c_n(\gamma)= (-1)^n \frac{\Gamma(\gamma + 1) }{\Gamma(n+1) \Gamma(\gamma-n+1)}
= \frac{\Gamma(n-\gamma)}{\Gamma(-\gamma) \Gamma(n+1)}.
\end{equation}
An immediate consequence of the periodicity is that
$\omega$ gets discretized, which is consistent with the 
appearance of discrete energy levels for a finite system. 
The integral over the exponentials in the expansion (\ref{Taylorexp})
will give delta-functions at those special values of $\omega$ and we
can try to extract an effective behavior in the prefactors, i.e.~the 
coefficients $c_n$. For a single channel case we can 
verify by inspection that the asymptotic behavior of the prefactor
in Eq.~(\ref{Taylorcoeff}) gives the expected power-law for large $n$,
i.e.~the semi-infinite length result can be recovered.

However, for two channels and arbitrary values of $\omega$ we need to
make a more careful analysis.  By using the multiplication formula 
\begin{equation} \label{prod.form.series}
\sum_{k=0}^{\infty}a_k x^k \cdot \sum_{k=0}^{\infty}b_k x^k =
\sum_{k=0}^{\infty}c_k x^k \ \ \ \ \ \ c_n=\sum_{k=0}^n a_k b_{n-k}
\end{equation}
we derive from (\ref{finiteLGF}), (\ref{N_omega}), (\ref{rep}), and 
(\ref{Taylorexp})
\begin{eqnarray} \label{NT0Lfinite}
N(\omega,r)&=& \frac{1}{\alpha \pi} 
\left(\frac{\pi \alpha}{L}\right)^{a_s+a_c}
\left(2 \sin{\frac{\pi r}{L}} \right)^{b_s+b_c} 
\sum_{n=0}^{\infty} \sum_{m=0}^{\infty} f_n (a_c,b_c,r) f_m (a_s,b_s,r)
\nonumber \\ & &  \times 
\left[\delta\left(\omega-\frac{\pi v_c (a_c +b_c)}{2L}-
\frac{\pi v_s (a_s + b_s)}{2L}
-n \frac{\pi v_c}{L}-m \frac{\pi v_s}{L}\right) 
 \right. \\ & &\hspace{3cm} \left.  
+\delta\left(\omega+\frac{\pi v_c (a_c+b_c)}{2L}
+\frac{\pi v_s (a_s+b_s)}{2L}+n 
\frac{\pi v_c}{L}+m \frac{\pi v_s}{L}\right) \right], \nonumber
\end{eqnarray}
where
\begin{equation}
f_n (k_1,k_2,r)=\sum_{p=0}^{n} c_{n-p}(-k_1)\sum_{q=0}^{p}c_{q}(-k_2/2)
c_{p-q}(-k_2/2) \cos{\left[(p-2q)\frac{2 \pi  r}{L}\right]} \ \ .
\end{equation}

\begin{figure}
\begin{center}
\mbox{\epsfxsize=4.3in \epsfbox{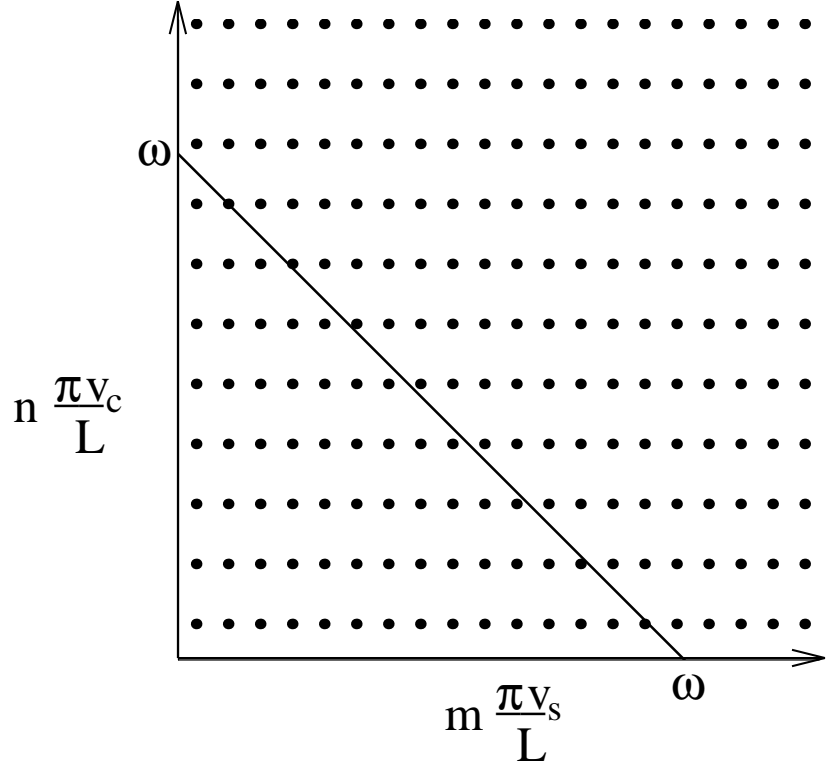}}
\end{center}
\caption{Illustration of  the allowed values of $\omega$ (points)
and the zeros of the argument of the delta-functions (line) in
Eq.~(\protect{\ref{NT0Lfinite}}).
The coordinates $n$ and $m$ are the summation indices. The line moves
outward as $\omega$ increases and a delta-function appears in the spectral
density at values of $\omega$ where the line
crosses a point with coefficients as
given in Eq. (\protect{\ref{NT0Lfinite}}). }
\label{sumsumdeltaill}
\end{figure}

To understand the role of the 
delta-functions in (\ref{NT0Lfinite}) it is convenient to represent the
argument of the delta function
in a two-dimensional parameter space, coordinatized by the pair of summation 
indices $(n,m)$ in (\ref{NT0Lfinite}), 
as shown in Fig.~(\ref{sumsumdeltaill}). 
The line connecting $(\omega,0)$ and $(0,\omega)$ indicates where the argument
of the delta-function vanishes, and hence selects the terms to be 
included in the double sum in (\ref{NT0Lfinite}).    The points are the
allowed values of $\omega$ and the ratio $v_s/v_c$ determines the relative
distances between the points in the x- and y-directions.
This ratio plays a crucial role, because when $v_s/v_c$ is a rational number
we have a resonance situation and the spectra will consist of
peaks with constant spacing. On the other hand, for $v_s/v_c$ irrational, 
$N(\omega,r)$ is still discrete for small $\omega$ but approaches a 
continuous function for large $\omega$, since the number of points
close to the line increases with increasing $\omega$. 
Moreover, if the spin-wave velocity $v_s$ is significantly smaller than
$v_c$ the spectrum may appear continuous, but the discrete
charge peaks may still
be resolvable. (As $v_s \to 0$ the charge excitations are described by
yet another exponent\cite{Mila2}.)

When the (experimental) energy resolution 
$\Delta \omega$ is larger than the spacing between the peaks,
it is appropriate to convert the infinite sums in Eq.~(\ref{NT0Lfinite}) into
integrations over continuous variables.  The resulting double integrals
can actually be done exactly in the extreme boundary case as well as in
 the extreme bulk case, and the semi-infinite length results are recovered
in both scenarios.  Even for the intermediate case it appears that 
the coefficients (\ref{Taylorcoeff}) under the double sum closely 
reproduce the power-laws of the semi-infinite case if some smearing
is taking into account. This is in strong contrast to the exponents that
describe the momentum distribution, which are known to be
strongly influenced by finite size effects as well as boundary 
effects\cite{OurPRL}.

In conclusion,
the main difference between a semi-infinite and a finite system is therefore
the appearance of a discretized spectrum, with possible beatings of
charge and spin excitations.  This effect can only be observed
in very small (mesoscopic) systems or with a very high experimental
resolution, since the smeared spectral weight appears to follow
the same frequency dependence.  The effect of the boundary remains dominant
for small frequencies in either case.

\subsection{The finite temperature and finite length limit}   

The most general case is to consider both finite temperature and a confined
system (finite length).  The periodicity of the Green's function 
(\ref{finiteLfiniteTGF}) is
unchanged, but we expect that the coefficients in front of the 
delta-functions will acquire temperature dependent corrections similar to 
the ones discussed in Sec.~\ref{TL0}.
We can make a similar Taylor-expansion as in the previous section by
using  Eq.~(\ref{prod.form.series}) to expand the temperature
dependent factor in Eq.~(\ref{boseexpval}) to the power of $\gamma$
\begin{equation}
\prod_{k=1}^\infty \left[1+\left(\frac{\sin\frac{\pi z}{2L}}
{\sinh{k\frac{\pi v_\nu \beta}{2L}}}\right)^2 \right]^{\gamma}=
\prod_{k=1}^\infty \sum_{n=0}^{\infty} d_n(\gamma,z) e^{-k n \beta 
\frac{\pi v_\nu}{L}}
\end{equation}
where
\begin{equation}
d_n(\gamma,z)=(-1)^n \sum_{p=0}^{n} c_{n-p}(-2 \gamma)\sum_{q=0}^{p}
c_{q}(\gamma)
c_{p-q}(\gamma) e^{i(p-2q)\frac{\pi z}{L}}, \ \ \label{coeff}
\end{equation}
and the coefficients $c_n$ are defined in Eq.~(\ref{Taylorcoeff}).
This can be written as
\begin{equation}
\prod_{k=1}^\infty \left[1+\left(\frac{\sin\frac{\pi z}{2L}}
{\sinh{k\frac{\pi v_\nu \beta}{2L}}}\right)^2 \right]^{\gamma}=
1+\sum_{m=1}^{\infty}  \sum_{n=0}^{m} g_{n,m}(\gamma) 
 \left( e^{i n \frac{\pi z}{L}} + e^{-i n \frac{\pi z}{L}}
\right)  e^{-m \beta \frac{\pi v_\nu}{L}} \label{g.nm}
\end{equation}
where $g_{n,m}(\gamma)$ is a highly nontrivial but well-behaved function 
composed of the coefficients $c_n(\gamma)$. The zero mode term $H$
also makes a non-trivial contribution which will be discussed 
elsewhere\cite{EMK}.

The integration over the exponentials in Eq.~(\ref{g.nm}) gives us
again delta functions.
However,  there is no shift or smearing of the 
peaks in the spectra due to temperature, only the
height of the existing peaks are modified, i.e.~the points where
the delta-functions contribute are at the same values of $\omega$ as indicated 
in Fig.~(\ref{sumsumdeltaill}). In the extreme boundary case we
can derive an explicit (but complicated) expression  
for the spectral density and we observe that the temperature
has a negligible effect for large $\omega$ as expected.  We  
conjecture quadratic behavior of the coefficients for small $\omega$,
which is supported by preliminary numerical evidence, i.e.~we observe a
similar effect to the one discussed in section \ref{TL0}. 
Thus, we are left with the analogous conclusion from the
previous section that finite size effects always result in a discrete
level spacing of delta-functions, but do not alter the (smeared) dependence on 
frequency.  We therefore recover the same cross-over from $\omega^2$-behavior
to boundary or bulk behavior as discussed in Sec.~\ref{TL0}.

\section{Discussion}

In conclusion, we have derived an exact closed-form expression for the 
single-electron Green's function of a spinful Luttinger liquid at finite 
temperature and confined to a finite interval by open boundaries. By 
analyzing the corresponding spectral density we have obtained detailed 
information about the interplay between boundary, finite-size and 
temperature effects in an interacting electron system. Most importantly, 
we find that the scaling of the zero-temperature spectral density with 
frequency $\omega$ close to the Fermi level is {\em always} governed by a  
coupling-dependent boundary exponent significantly larger than the bulk 
exponent.  In other words, the 
asymptotic low-energy behavior of a Luttinger liquid with an open 
boundary belongs to a different universality class than that of the bulk 
theory. Thermal fluctuations at finite temperature $T>0$ destroy 
this behavior and 
open up a new regime for $\omega$ less than the temperature where the spectral 
density exhibits quadratic scaling in $\omega$. Not surprisingly, the {\em 
same} effect is present in a bulk system for this frequency 
range\cite{nakamura},
implying that the boundary plays no decisive role in the process. In the 
case of a finite interval confined by {\em two} open boundaries, our results
reveal  a discretized spectrum with delta-functions at the allowed
energy levels.  It is interesting to observe the 
dependence of spacing of the energy levels on the ratio 
$v_c/v_s$ between the effective velocities $v_c$ and $v_s$ of the charge- 
and spin excitations, respectively: For $v_c/v_s$ a rational number, the 
spectrum consists of well-separated peaks which, for sufficiently large 
$\omega$, coalesce to a quasi-continuum if $v_c/v_s$ is shifted to an 
irrational number.
Although this effect can not be observed experimentally, it nonetheless 
suggests a resonance phenomenon with the collective charge and spin 
excitations showing interference effects at special values of the
electron-electron coupling. 
However, despite the appearance of a discrete spectrum we find that 
finite size does not influence the $\omega$ dependence of the
integrated (i.e.~smeared) spectral 
density significantly, so that the same boundary and finite
temperature effects as for a semi-infinite can be observed.

Let us close by briefly discussing the possible relevance of our results 
to experiments, in particular the photoemission studies on  
the Bechgaard salts\cite{DardelHwu} referred to in the introduction. 
These materials are composed of molecular chains which become conducting 
above some characteristic temperature $T_c$, and are 
expected to show Luttinger liquid behavior provided the temperature is 
high enough to mask the weak inter-chain coupling. However, as mentioned 
in the introduction, high-precision photoemission experiments indicate a 
scaling of spectral weight with frequency that is
inconsistent with standard theory of a bulk Luttinger liquid: 
The effective exponent $\alpha$ for scaling of 
the photoemission intensity is roughly 1.25 (Ref. \onlinecite{DardelHwu})
as also seen in independent 
NMR experiments on the same materials \cite{Wzietek}, 
whereas the largest realistic 
value obtainable from a {\it bulk} Luttinger liquid description is $\alpha = 
0.125$, 
corresponding to the large-U Hubbard chain \cite{Voit}. Attempts to 
include long-range Coulomb repulsion, which can be shown to increase 
$\alpha$ \cite{Schulz}, fails due to the instability against an 
insulating phase at 
$\alpha = 9/16$ \cite{Mila}, so other explanations must be invoked. As 
the typical escape depth of photoelectrons in the UV range is only 5 - 10 
\AA, the experiments are extremely surface sensitive, suggesting that
1D boundary effects may play a role for the observed scaling behavior.

Consider first the case where one probes electrons that escape from a 
crystal face perpendicular to the 1D molecular chains. The photoemission 
intensity $I(\omega,\beta)$ is then proportional to the local spectral density 
$N(\omega, \beta, r)$, integrated over the escape depth of the 
photoelectrons, and weighted by the Fermi-Dirac distribution 
$f_{FD}(\beta \omega)$
\begin{equation}
I(\omega,\beta) \propto 
\int dr \ f_{FD}(\beta \omega) N(\omega, \beta, r) \ .
\label{intensity}
\end{equation}
In a boundary dominated region, $I(\omega)$ is seen to be dramatically 
reduced compared to the a bulk regime, considering our results in 
Eq.~(\ref{Nextreme}). With a typical escape depth of a few lattice spacings, 
the condition for boundary behavior $r\omega/v_c < 1$ may apply over an 
energy range of several hundred meV (since $v_c > a E_F$, with $a$ 
the lattice spacing and with $E_F \approx 0.5 - 1$ eV, depending on the 
particular material \cite{JeromeSchulz}).

In the recent photoemission experiments on $\rm (TMTSF)_{2}PF_6$ 
(Ref. \onlinecite{DardelHwu}), 
the chains are always in the plane of the cleaved 
surface \cite{Grioni}, and it is less clear to what extent 1D boundary 
effects contribute. However, in the likely case that the cleaving of the 
surface introduces defects in neighboring chains, effectively breaking these 
into smaller segments, we may model the breaks by open boundaries 
and apply our results. 
Unfortunately, the actual defect concentration  
remains unknown, and it is therefore
difficult to make a quantitative prediction from our results. This is 
an important issue that in principle should be possible to resolve via 
STM techniques. To explore the size of the 
boundary effects experimentally, it
would also be of great interest to do photoemission experiments
on cleaved surfaces that are {\it perpendicular} to the 
chains that could then be compared to the results from cleaved surfaces
parallel to the chains.

\begin{figure}
\begin{center}
\mbox{\epsfxsize=6.3in \epsfbox{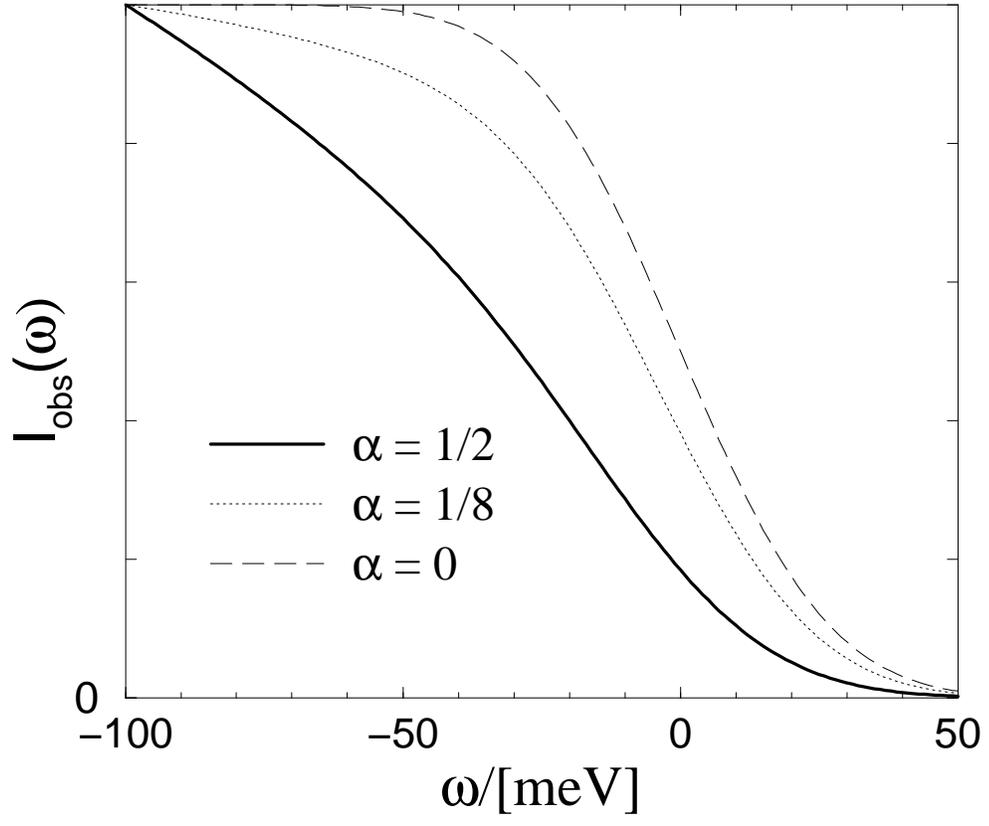}}
\end{center}
\caption{The predicted intensity $I_{\rm obs}$ in arbitrary units as a
function of $\omega$
for boundary and bulk cases (i.e.~for power-laws with $\alpha_{\rm bound}=1/2$ 
and $\alpha_{\rm bulk} = 1/8$, respectively). 
The corresponding three-dimensional case ($\alpha = 0$) is also shown.
Temperature ($T=50K$) and finite resolution ($\Delta = 20meV$) effects have
been taken into account. }
\label{exp}
\end{figure}

As discussed in \cite{OurPRL}, the finite-energy 
resolution of the photon lines effectively introduces an 
averaging over the ``true'' spectrum
\begin{equation}
I(\omega)_{\rm obs} \equiv \frac{1}{\sqrt{2 \pi} \Delta} \int 
e^{-(\omega-x)^2/2 \Delta^2} I(x) \ dx\label{GaussAverage}
\end{equation}
which completely wipes out the power-law singularities in either the 
bulk- or boundary case. This ``smearing'' results in
 a similar effect as the thermal fluctuations, which also wipe out the 
sharp cusp from the power-laws as shown in Fig.~(\ref{n_ro_roverbeta}),
but have to be taken into account separately.
With an experimental resolution of $\Delta = 20 
meV$ and at a temperature $T=50K$ (experimental values according to 
\cite{DardelHwu}), 
and {\em assuming} boundary dominated behavior for $N(\omega, \beta, r)$, the 
observed intensity in the vicinity of the Fermi level indeed appears to 
be depleted with an exponent of one or larger as shown in Fig. (\ref{exp}).
 (This should be compared to the large-U 
Hubbard exponent $\alpha_{\rm bulk} = 1/8$ of the bulk spectral 
function without 
temperature or averaging effects). In experiments the condition for 
boundary behavior $\omega < v_c/r$ will be satisfied over an energy range 
$\omega \sim E_F/r$ around the Fermi-level, where $r$ is the 
distance from the boundary in 
units of the lattice spacing. This means that if the broken chains close to 
the cleaved surface have an average 
impurity density of a few percent, boundary effects could be 
observed over a region of up to $100 meV$ around the Fermi energy. 
Experiments indeed suggest a scaling $I(\omega)_{obs} \propto 
\omega^{\alpha_{obs}}$ with $\alpha_{obs} > 1$, extending, however, over 
a larger energy range. Thus, some additional mechanism (inter-chain 
coupling or electron-phonon coupling \cite{voitschulz}) 
most likely have to be invoked to fully explain the data. Yet, a complete 
modeling of photoelectron spectroscopy on 
quasi-1D organic metals must certainly 
incorporate the boundary and temperature effects predicted in the present 
paper.

\begin{acknowledgements}
We thank  M. Grioni and J. Kinaret for valuable correspondence.
This work was supported in part by the
Swedish Natural Science Research Council.
\end{acknowledgements} 

\appendix
\section{Calculation of the chiral Green's function}
\label{App.Green}
Here we extend the calculations of Ref. \onlinecite{EMK} to the case
of spinful Fermions with open boundaries.
To calculate the chiral fermionic Green's function, we find it useful
to treat the contributions from the zero modes and the
dynamic bosonic modes separately.  Hence we write $\phi_{\nu,L}$ in 
Eq.~(\ref{eqn:Lmodeexp}) as a sum of the zero modes and the 
harmonic oscillator terms 
\begin{equation}
\phi_{\nu,L}\ =\ \frac{1}{2}\left( \phi_{\nu,0}+\tilde{\phi}_{\nu,0} \right)
+\hat{Q}_\nu \frac{x+v_{\nu}t}{2L}+ S_{\nu,L},
\end{equation}
where the bosonic operators are contained in the sum
\begin{equation}
\label{eqn:Smodeexp}
S_{\nu,L}=\sum_{n=1}^\infty \frac{1}{\sqrt{n \pi}}
\left[ e^{- i \frac{n \pi \left(x+v_{\nu}t\right)}{L}} a_n^{\nu}
+ \mbox{h.c.} \right] \, .
\end{equation}
We now insert this mode expansion into
the bosonization formula  (\ref{eqn:ourbosform}) for $\psi_L$, and 
use the definition of $G(x,y,t) \equiv \left<\psi^{\dagger}_{L,\sigma}(x,t) 
\psi^{}_{L,\sigma}(y,0) \right>$ to find
\begin{eqnarray}\label{Gapp}
G(x,y,t) \ \propto&& H(x,y,t)
\prod_{\nu = c,s} \exp\left[-\frac{i \pi}{4 L} 
\left(v_\nu t + x-y\right)\right]  \nonumber \\
&&\times \exp\left[\frac{\pi}{2}
(K_\nu+K_\nu^{-1})^2 B_{\nu,L}(x,t;y,0)\right]
\exp\left[\frac{\pi}{2}(K_\nu-K_\nu^{-1})^2 B_{\nu,L}(-x,t;-y,0)\right]
\nonumber \\
&&\times\exp\left[\frac{\pi}{4}(K_\nu^{-2}-K_\nu^{2})
\left(2 B_{\nu,L}(x,t;-y,0)+2 B_{\nu,L}(-x,t;y,0)\phantom{\frac{}{}}
\right.\right.\\
&&\hspace{.2cm}\left. \left.\phantom{\frac{}{}}
-B_{\nu,L}(x,t;-x,t)- B_{\nu,L}(-x,t;x,t)
-B_{\nu,L}(y,0;-y,0)-B_{\nu,L}(-y,0;y,0) \right)\right]. \nonumber
\end{eqnarray}
Here we used the identity
$e^A e^B = : e^{A+B} : e^{\left< AB+\frac{A^2+B^2}{2}\right>}$ 
for the bosonic operators
$S_{\nu,L}$ and we have defined the bosonic Green's function
\begin{eqnarray} \label{bosonGF}
B_{\nu,L}(x,t;x',t')&=&\left<S_{\nu,L} (x,t)S_{\nu,L} (x',t')-
\frac{1}{2}\left[S_{\nu,L} (x,t)S_{\nu,L} (x,t)+
S_{\nu,L} (x',t')S_{\nu,L} (x',t')\right]\right>.
\end{eqnarray}
The contribution from the zero modes is
\begin{eqnarray}
H(x,y,t)&=&\left< \prod_{\nu = c,s}\exp iu_\nu 
\sqrt{\frac{2}{\pi}}K_\nu \hat{Q}_\nu\right> 
\label{zeromodepart}
\end{eqnarray}
where $u_\nu=-\frac{\pi}{2L}\left(v_\nu  K_\nu^{-2}t + x-y\right)$.
This factor {\it cannot} be written as a product of spin and charge 
expectation values separately, because the quantum numbers $n$ and $m$
in Eq.~(\ref{nontrivial})
are connected by the condition that both are even or both are odd.
However, since we know the quantization condition (\ref{nontrivial}) and
the energy spectrum (\ref{spectrum}) for the zero modes, we can
directly sum over all eigenvalues $m,n$.  The factor $H$ can then be expressed
in terms of the elliptic theta functions (see sections 8.18-8.19 in~\cite{GR})
\begin{eqnarray}
H(x,y,t)&=&
\frac{\vartheta_2(u_c+ \tau_c k_F L| \tau_c)
\vartheta_3(u_s|\tau_s)+\vartheta_3(u_c+ \tau_c k_F L| \tau_c)
\vartheta_2(u_s|\tau_s) }{\vartheta_2( \tau_c k_F L| \tau_c)
\vartheta_3(0|\tau_s)+\vartheta_3( \tau_c k_F L| \tau_c)
\vartheta_2(0|\tau_s)}e^{i 2 u_c \frac{k_F L}{\pi}},
\end{eqnarray}
where $\tau_\nu = i \frac{v_\nu \beta}{K_\nu^2 L}$.

To calculate the bosonic Green's function $B_{\nu,L}$,
we insert the expression (\ref{eqn:Smodeexp}) for $S_{\nu,L} (x,t)$
into Eq.~(\ref{bosonGF}), which gives
\begin{eqnarray}
\label{eqn:app:step1}
B_{\nu,L}&=&
\sum_{n=1}^{\infty} \frac{1}{4\pi n} \left[
\left( e^{-2\pi i n \frac{v_\nu t +x - (v_\nu t' +x')}{2L}}-1\right)
\left(1+m_n^\nu\right)+
\left( e^{2\pi i n \frac{v_\nu t +x - (v_\nu t' +x')}{2L}}-1\right)
m_n^\nu\right],
\end{eqnarray}
where $m_n^\nu$ are the Bose-Einstein distributions
\begin{equation}
m_n^\nu=\left<{a_n^\nu}^\dagger {a_n^\nu}^{}\right>=
\frac{1}{e^{\beta \frac{n v_\nu \pi}{L}}-1}.
\end{equation}
We define
$a=e^{-2\pi i \frac{v_\nu t +x - (v_\nu t' +x')}{2L}}$ and
$b=e^{\beta \frac{v_\nu \pi}{L}}$, which allows us to write
\begin{equation}
\frac{1}{1-(b^{-1})^n}=\sum_{k=0}^\infty(b^{-n})^k=\sum_{k=0}^\infty(b^{-k})^n.
\end{equation}
The bosonic Green's function can then be written as
\begin{eqnarray} \label{sum}
B_{\nu,L}&=& 
\sum_{n=1}^{\infty} \frac{1}{4\pi n} \left[
\left( a^n-1\right) \sum_{k=0}^\infty\left(b^{-k}\right)^n
+\left( a^{-n}-1\right) b^{-n} \sum_{k=0}^\infty\left(b^{-k}\right)^n \right]
\nonumber \\ &=&
\sum_{k=0}^\infty\sum_{n=1}^{\infty}\frac{1}{4\pi n}\left[\phantom{\frac{}{}}
\left(a b^{-k}\right)^n - \left(b^{-k}\right)^n + \left(a^{-1} b^{-1-k}
\right)^n - \left(b^{-1-k}\right)^n \phantom{\frac{}{}} \right]
\\ &=&
\sum_{n=1}^\infty \frac{1}{4\pi n}\left( a^n-1\right) +
\sum_{k=1}^\infty\sum_{n=1}^{\infty}\frac{1}{4\pi n}\left[\phantom{\frac{}{}}
\left(a b^{-k}\right)^n - \left(b^{-k}\right)^n + \left(a^{-1} b^{-k}\right)^n 
- \left(b^{-k}\right)^n \phantom{\frac{}{}} \right].\nonumber
\end{eqnarray}
We now can use the formula
\begin{equation}
\sum_{k=1}^\infty \frac{z^k}{k} = - \ln (1-z), \ \ |z|<1
\end{equation}
and we need to use the high momentum cut-off $\alpha$ in 
the first two terms of Eq.~(\ref{sum})
\begin{equation}
\sum_{n=1}^\infty \frac{1}{n}\left( a^n-1\right)
\longrightarrow \sum_{n=1}^\infty \frac{1}{n}\left( a^n-1\right)c^n, \ \ \ \ 
c=e^{-\alpha \frac{\pi}{L}}, \ \ \ \
\alpha \rightarrow 0.
\end{equation}
This yields
\begin{eqnarray} \label{Sab}
B_{\nu,L}&=&-\frac{1}{4\pi}\ln \left[\frac{1-ac}{1-c}\prod_{k=1}^\infty
\frac{(1-a b^{-k})(1-a^{-1} b^{-k})}{(1-b^{-k})(1-b^{-k})}\right] \nonumber \\
&=&-\frac{1}{4\pi}\ln \left[\frac{(ac)^{1/2}}{\frac{-i \alpha \pi}{2 L}}
\left(\frac{(ac)^{-1/2}-(ac)^{1/2}}{2i}\right) \prod_{k=1}^\infty
\left(1+\frac{\left(\frac{a^{-1/2}-a^{1/2}}{2i}\right)^2}
{\left(\frac{b^{k/2}-b^{-k/2}}{2}\right)^2}\right) \right].
\end{eqnarray}
Inserting $a$ and $b$ defined above gives us
\begin{eqnarray}
B_{\nu,L}(x,t;x',t')&=&
\frac{i}{4} \left( \frac{v_\nu t +x - v_\nu t' -x'}{2L}\right)
-\frac{1}{4\pi} \ln \left[\phantom{\frac{}{}}
F_\nu(v_\nu t +x - v_\nu t' -x')\phantom{\frac{}{}} \right] \label{BnL}
\end{eqnarray}
where 
\begin{equation}
F_\nu(z)=i {\frac{2L}{\alpha \pi} \sin\frac{\pi z}{2L}}
\prod_{k=1}^\infty \left[1+\left(\frac{\sin\frac{\pi z}{2L}}
{\sinh{k\frac{\pi v_\nu \beta}{2L}}}\right)^2 \right] \ = \
\frac{\vartheta_1(\frac{\pi z}{2 L}|i\frac{v_\nu \beta}{2 L})}{\vartheta_1(
-i\frac{\pi \alpha}{2 L}|i\frac{v_\nu \beta}{2 L})}\, .
\end{equation}
 The first term in
(\ref{BnL}) cancels with the phase in the zero mode part of
(\ref{Gapp}), resulting in Eq.~(\ref{finiteLfiniteTGF}).

 
\end{document}